\documentclass[aps,prd,preprintnumbers,nofootinbib,floatfix,twocolumn]{revtex4}

\usepackage{bm}
\usepackage{latexsym}
\usepackage{dcolumn}
\usepackage{amsfonts,amssymb}
\usepackage{graphicx,epsfig}

\def\eq#1{{Eq.~(\ref{#1})}}

\begin{document}


\title{Einstein's equations as a thermodynamic identity: The cases of stationary axisymmetric horizons and evolving spherically symmetric horizons}
\author{Dawood Kothawala}
\email[1]{dawood@iucaa.ernet.in}
\author{Sudipta Sarkar}
\email[2]{sudipta@iucaa.ernet.in}
\author{T. Padmanabhan}
\email[3]{paddy@iucaa.ernet.in}
\affiliation{IUCAA,
Post Bag 4, Ganeshkhind, Pune - 411 007, India\\}
\date{\today}


\begin{abstract}
There is an intriguing analogy between the gravitational dynamics of the horizons and thermodynamics. In case of general relativity, as well as for a wider class of Lanczos-Lovelock theories of gravity, it is possible to interpret the field equations near any spherically symmetric horizon as a thermodynamic identity $TdS = dE + PdV$.  We study this approach further and generalize the results to two more generic cases within the context of general relativity: (i) stationary axis-symmetric horizons and (ii) time dependent evolving horizons . In both the cases, the near horizon structure of Einstein equations can be expressed as a thermodynamic identity under the virtual displacement of the horizon. This result demonstrates the fact that the thermodynamic interpretation of gravitational dynamics is \textit{not} restricted to spherically symmetric or static horizons but is quite generic in nature and indicates a deeper connection between gravity and thermodynamics.
\end{abstract}

\maketitle
\vskip 0.5 in
\noindent
\maketitle
\section{INTRODUCTION}
Spacetimes with horizons show an interesting resemblance to thermodynamic systems with well defined notions of temperature and entropy. General Relativity (GR) allows the existence of spacetime horizons which act as a casual boundary and block the propagation of any information to an outside observer. This led Bekenstein \cite{bek} in 1973 to argue that black holes must posses a non-zero entropy since they hide information from the outside observer. The picture became clearer in 1975, when Hawking showed \cite{hawkings} that one can attribute a  temperature $k_B T = \hbar c/8 \pi M$ to a black hole of mass $M$. This suggested the possibility  that the black holes behave like thermal systems and the laws of black hole mechanics are basically same as the laws of thermodynamics \cite{blackthermo}.

It was soon realized that one can attribute a temperature to several other types of horizons thereby suggesting a generic connection between thermodynamics of horizons and of gravity. This connection, however, is not yet understood at a deeper level \cite{paddy1}.
One possible
paradigm envisages gravity as an emergent phenomenon analogous to the theory of
elasticity of a deformable solid. Then the microscopic degrees of
freedom of spacetime (analogous to the atoms in the
case of solids) will play a role only when spacetime is probed at
Planck scales (which would be analogous to the lattice spacing of a 
solid \cite{zeropoint}).
However, in a manner
which is not fully understood, the horizons --- 
which block information from certain classes of observers ---
link \cite{magglass} 
aspects of microscopic physics with  the bulk dynamics just as
thermodynamics can provide a link between statistical mechanics and
(zero temperature) dynamics of a solid. 
If this picture is correct, one should be able to connect the field equations describing the dynamics of gravity with the horizon thermodynamics. 

There have been several approaches which have attempted to do this with different levels of success \cite{sakharov,paddy1,paddyholo}. The most explicit demonstration of this fact occurs in the case of spherically symmetric horizons in Einstein-Hilbert gravity \cite{paddy2}. In that case, the near horizon structure of the field equation can be cast as a thermodynamic identity $TdS= dE + PdV$ arising from the virtual displacement of the horizon normal to itself. Recently this result has been extended to the spherically symmetric horizons in Lanczos-Lovelock gravity \cite{Lovelock} and it has been shown in that case also that the field equations can be interpreted as a thermodynamic relation. Since the extra, higher derivative terms in Lanczos-Lovelock gravity can be thought of as quantum corrections to GR, this non-trivial result suggests the possibility that the thermodynamic interpretation of the field equations remains valid even if one includes possible quantum corrections to Einstein-Hilbert action functional motivated by quantum gravity models.
Further, it is possible to show that the Einstein equations evaluated on the apparent horizon is a thermodynamic identity \cite{cai} even in the case of Friedmann universe.
Some other recent work attempts to provide the structural reasons for the existence of these results \cite{ayan,aseem2}.

So far, all these calculations have been performed in a spherically symmetric and time independent setting. Assumption of spherical symmetry allows us to write the metric in a simple form. It is then  possible to find a  Rindler limit to this metric near the horizon and  the horizon surface being spherical, the near horizon structure of the field equations becomes simple even in the case of general Lovelock gravity. Similarly, the assumption of time independence restricts the analysis only to static horizons. If the thermodynamic interpretation of gravity is to be generic, one needs to go beyond these assumptions and check whether the results remain valid for more general spacetimes. Intuitively we expect these results to hold more generally because one of the crucial inputs (arising from spherical symmetry) is the equality of $T_{t}^{t}$ and $T_{r}^{r}$ on the horizon and it has been shown that this result is valid even for general stationary killing horizons \cite{medvid}. So a step forward would be to study the near horizon structure of the field equations for a more general spacetimes.

In this paper we consider two such generalizations taking one at a time. First, we generalize the results for a time dependent but spherically symmetric metric in which the horizon is evolving. Next we consider the Kerr-Newman solution which describes a non spherically geometry (but time independent). For the first case the horizon is evolving in time and there is no timelike killing vector and in the second the intrinsic geometry of the  horizon is not a $2$-sphere; these two cases, therefore, generalizes the previous results in two different directions. Remarkably, \textit{even in these two cases, the field equations near the horizon can be written as a thermodynamic relation}. This result indicates that the thermodynamic interpretation of the field equations is not restricted to the static and spherically symmetric case but also applicable to the evolving horizons and stationary axis-symmetric spacetimes at least in the context of general relativity.

The organization of the paper is as follows: in the next section we will briefly review the case of static spherically symmetric horizons. In section III, we will present the analysis for evolving horizons. In section IV we will consider the Kerr-Newman spacetime and finally discuss the implications in section V.

\section{Warm up: spherically symmetric Static horizons}

Consider a static, spherically symmetric horizon, in a spacetime described by a metric:
\begin{equation}
ds^2 = -f(r) c^2 dt^2 + \frac{1}{g(r)} dr^2 + r^2 d\Omega^2. \label{spmetric}
\end{equation}
We will assume that the horizon is given by a simple zero of the function $f(r)$ at $r=a$. The temperature associated with this horizon is $k_BT=\hbar c \sqrt{f'(a)g'(a)}/4\pi$ where we have introduced normal units. (Even for spacetimes with multi-horizons this prescription is locally valid for each horizon surface.).
We, however, need to ensure that the surface $r=a$ is just a null surface and not a singularity. This requires two conditions on $g(r)$ at $r=a$. (i) We need $g(a)=0$ and
(ii) $f'(a)=g'(a)$ where the second condition ensures the the regularity of the Ricci curvature on the horizon. Because of this, the temperature associated with the horizon at $r=a$ becomes $k_BT=\hbar c g'(a)/4\pi$. Therefore under these conditions, the energy momentum tensor on the horizon must have the form,
\begin{eqnarray}
T_{t}^{t}|_{r=a} = T_{r}^{r}|_{r=a} ; ~~~T_{\theta}^{\theta}|_{r=a} = T_{\phi}^{\phi}|_{r=a}. \label{emten}
\end{eqnarray}

Next consider the Einstein equation for this metric, given by
$(1-g)-rg'(r)=-(8\pi G/c^4) Pr^2$ (where $P = T^{r}_{r}$ is the radial pressure) and evaluate it at $r=a$. This gives:
\begin{equation}
\frac{c^4}{G}\left[{1\over 2} g'(a)a - \frac{1}{2}\right] = 4\pi P a^2
\label{reqa}
\end{equation}
If we now consider two solutions with two different radii $a$ and $a+da$ for the horizon,
then multiplying the \eq{reqa} by $da$, and introducing a $\hbar$ factor \textit{by hand} into an otherwise classical equation, we can rewrite it as
\begin{equation}
   \underbrace{\frac{{{\hbar}} cg'(a)}{4\pi}}_{\displaystyle{k_BT}}
    \ \underbrace{\frac{c^3}{G{{\hbar}}}d\left( {1\over 4} 4\pi a^2 \right)}_{
    \displaystyle{dS}}
  \ \underbrace{-\ {1\over 2}\frac{c^4 da}{G}}_{
    \displaystyle{-dE}}
 = \underbrace{P d \left( {4\pi \over 3}  a^3 \right)  }_{
    \displaystyle{P\, dV}}
\label{EHthermo}
\end{equation}
and read off the expressions:
\begin{equation}
 S={1\over 4L_P^2} (4\pi a^2) = {1\over 4} {A_H\over L_P^2}; \quad E={c^4\over 2G} a
    =\frac{c^4}{G}\left( {A_H\over 16 \pi}\right)^{1/2}
\end{equation}
where $A_H$ is the horizon area and $L_P^2=G\hbar/c^3$.
Some comments are relevant regarding this result, especially since they are valid for our generalization discussed later as well:

(a) The combination $TdS$ is completely classical and is independent of $\hbar$ but $T\propto \hbar$ and $S\propto 1/\hbar$. This is analogous to the situation in classical thermodynamics when compared to statistical mechanics. The $TdS$ in thermodynamics is independent of Boltzmann's constant while statistical mechanics will lead to an $S\propto k_B$ and $T\propto1/k_B$.

(b) In spite of superficial similarity, \eq{EHthermo} is different
from the conventional first law of black hole thermodynamics, (as well
as some previous attempts to relate thermodynamics and gravity, like
e.g. the second paper in ref. \cite{sakharov}), due to the presence of
$PdV$ term. This relation  is more in tune with the membrane paradigm
\cite{membrane} for the blackholes. This is easily seen, for example,
in the case of Reissner-Nordstrom blackhole for which $P\neq0$. If a
\textit{chargeless} particle of mass $dM$ is dropped into a
Reissner-Nordstrom blackhole, then an elementary calculation shows
that the energy defined above as $E\equiv a/2$ changes by $dE= (da/2)
=(1/2)[a/(a-M)]dM\neq dM$ while it is $dE+PdV$ which is precisely
equal to $dM$ making sure $TdS=dM$. So we need the $PdV$ term to get
$TdS=dM$ when a \textit{chargeless} particle is dropped into a
Reissner-Nordstrom blackhole. More generally, if $da$ arises due to
changes $dM$ and $dQ$, it is easy to show  that \eq{EHthermo} gives
$TdS=dM -(Q/a)dQ$ where the second term arises from the electrostatic
contribution from the horizon surface charge as expected in the
membrane paradigm. 

(c) This result can be formally interpreted by noting that in standard thermodynamics, we consider two equilibrium states of a system differing infinitesimally in the extensive variables like entropy, energy, and volume by $dS$, $dE$ and $dV$ while having same values for the intensive variables like temperature $(T)$. Then, the first law of thermodynamics asserts that $TdS = PdV + dE$ for these states. In a similar way, Eq.~(\ref{EHthermo}) can be interpreted as a connection between two quasi-static equilibrium states where both of them are spherically symmetric solutions of Einstein equations with the radius of horizon differing by $da$ while having same source $T_{ij}$ and temperature $k_BT=\hbar c g'(a)/4\pi$. This formalism does not care what causes the change of the horizon radius and therefore much more formal and generally applicable.
Note that the structure of the equation itself allows us to ``read off" the expressions for entropy and energy.

The validity of this approach as well as the uniqueness of the results are discussed at length in ref. \cite{paddy2} and will not be repeated here.
Henceafter, we shall adopt natural units, in which $\hbar = c = G = 1$, the Boltzmann constant $k_B$ is also set to unity.
\section{Evolving horizons}
In order to study the case where the horizon evolves with time, we shall adopt the approach of ref. \cite{visser}, in which a simple definition of a time-dependent, spherically symmetric horizon has been introduced for a dynamical black hole. We will study a generalized version of the analysis given in \cite{visser}, and analyze a spherically symmetric time-dependent spacetime described by the metric,
\begin{eqnarray}
ds^2 &=& -f(t,r)~dt^2 + \frac{1}{g(t,r)}~dr^2 + r^2 d\Omega^2
\label{evolving1}
\end{eqnarray}
We will define the \textit{evolving horizon} at $r=r_H$ by the condition, $f(t,r_H)=0$. In principle this relationship can be inverted to determine $r_H(t)$. Again, the regularity of the Einstein equations and Ricci scalar on the horizon enforces several restrictions on the functions $f(t,r_H)$ and $g(t,r_H)$. In particular, we must have $ f(t,r_H)=g(t,r_H)=0$ and $f'(t,r_H)=g'(t,r_H)$. We need to associate a time dependent temperature with this evolving evolving horizon, for which we need the surface gravity, $\kappa$ associated with it. To determine $\kappa$, we will follow the approach in \cite{visser}, and write down the metric \eq{evolving1} in the Painleve-Gullstrand form,
\begin{eqnarray}
ds^2 &=& -[ c^2(\tau,r)-v^2(\tau,r) ]~d\tau^2 + 2~v(r,\tau)~dr~d\tau \\
&+& dr^2 + r^2 d\Omega^2,
\end{eqnarray}
where,
\begin{eqnarray}
c(\tau,r) = \frac{1}{\dot{\tau} } \sqrt{\frac{f(\tau,r)}{g(\tau,r)}}~~\mathrm{and}~~\\
v(\tau,r) = c(\tau,r) \sqrt{ 1 - g(\tau,r) },
\end{eqnarray}
where the prime denotes the derivative with respect to $r$ and dot is that with respect to $t$. In order to determine the surface gravity, we define the outward radial null vector as
\begin{eqnarray}
l^{\alpha} = \frac{ ( 1, c(\tau,r)-v(\tau,r), 0, 0 ) }{ c(\tau,r) },
\end{eqnarray}
and verify that $ g_{\alpha \beta} l^{\alpha} l^{\beta} = 0$. Because of spherical symmetry we must have,
\begin{eqnarray}
l^{\alpha} \nabla_{\alpha}l^{\beta} = \kappa_{l}l^{\beta},
\end{eqnarray}
where the scalar $\kappa_{l}$ is defined everywhere in the spacetime and should reduce to surface gravity $\kappa$ on the horizon. Evaluating $\kappa_{l}$ and putting the condition of the horizon the surface gravity is found to be,
\begin{eqnarray}
\kappa = \frac{g'(\tau,r_H)}{2}
\end{eqnarray}
It is easy to verify that this result has all the correct limits when one consider time independence. Although there is no  explicit time derivative in the expression of $\kappa$ but this result is quite in tune with the result obtained in \cite{visser} for a time dependent Schwarzschild like metric; although $\kappa$ does not involve any explicit time derivative  it is still dynamic in a sense that the horizon radius $r_H(\tau)$ is itself changing with time.
With this definition of the surface gravity $\kappa$, we can write the temperature associated with the evolving horizon as, $T = g'(t,r_{H}(t))/4 \pi$. The entropy of this evolving horizon is taken as one-fourth of the instantaneous area. 

The Einstein equation for the metric \eq{evolving1} evaluated on the horizon is,
\begin{eqnarray}
G^{t}_{t} = G^{r}_{r} = \frac{g'(r_H(t),t)}{r_{H}(t)} - \frac{1}{r_{H}^{2}(t)} = 8 \pi T_{r}^{r}.
\label{evolving2}
\end{eqnarray}
We now consider two solutions, with horizon radii $r_H$ and $r_H + dr_{H} $, where $dr_{H} = \dot{r_{H}}~dt$. Then, multiplying the above equation on both sides by $dr_{H}(t)$, and using the expression for the horizon temperature as obtained above, it is easily seen that the Einstein equations \eq{evolving2}, on the horizon, can be written as
\begin{eqnarray}
TdS[r_H(t)] - dE[t,r_H(t)] = P dV \label{thermoevol},
\end{eqnarray}
where $P=T^r_r$ is the radial pressure, $S[r_H(t)]=\pi r_H^2$ is the entropy and $dV=4 \pi r_{H}^2(t)~dr_{H}$ is the areal volume. We can now read off  $r_{H}(t)/2$ as the instantaneous Misner-Sharp energy $E(t)$ of the horizon. We conclude that for the evolving horizon also the near horizon structure of the Einstein equation can be interpreted as a thermodynamic identity $TdS = dE + PdV$ where the differentials arise due to the evolution of the horizon with time.

\section{Kerr-Newman spacetime}
The Kerr-Newman spacetime is a stationary axis-symmetric solution of the Einstein's equations in the presence of the Maxwell field. The form of the metric in Boyer-Lindquist coordinates $(t, r, \theta, \phi)$ is given by \cite{MTW},
\begin{eqnarray}
ds^2 &=&-\frac{\Delta ^2}{\rho ^2}\left( dt- a \sin^2 \theta~ d\phi
\right) ^2+\frac{\rho ^2}{ \Delta^2 }dr^2  \nonumber \\
&&\ + \rho ^2 d\theta ^2+\frac{ \sin
^2\theta}{\rho ^2}\left( a dt-(r^2+a^2) d\phi \right) ^2,
\label{kerrN}
\end{eqnarray}
where
\begin{eqnarray}
\Delta ^2 &=&(r^2+a^2) - 2Mr + Q^2,  \nonumber \\
\rho ^2 &=&r^2+a^2\cos ^2\theta .
\end{eqnarray}
and the vector potential $A_\mu $ is:
\begin{eqnarray}
A_\mu = -\frac{Q r}{\rho ^2}(\delta _\mu ^t-a \sin ^2\theta~  \delta
_\mu ^\phi ).  \label{Pot}
\end{eqnarray}
The metric has a true ring singularity at $ \rho(r , \theta) = 0 $. The outer event horizon is a null stationary 2-surface defined by $r = R = M + \sqrt{M^2 - a^2 - Q^2}$. The vector $\xi^{\alpha} = t^{\alpha} + \Omega_{H} \phi^{\alpha}$ is a killing vector of this spacetime and it is null on the horizon at $r = R$, where the quantity $\Omega_{H}$ is interpreted as the angular velocity on the horizon given by \cite{MTW},
\begin{eqnarray}
\Omega_{H} = \frac{a}{R^2 + a^2}.
\end{eqnarray}
Our aim is to study the near horizon structure of the field equations for this metric and verify the existence of  a possible thermodynamic interpretation. In case of spherical geometry one can write a wide class of spacetimes (like Schwarzschild, De-Sitter, Reissner-Nordstrom etc) in the form of \eq{spmetric}. There is no such general prescription available for a general stationary spacetime. Therefore, we have to use an ansatz to write a general stationary metric and extract the near horizon structure of the field equations. In order to achieve this, we replace $\Delta ^2 = (r^2+a^2) - 2Mr + Q^2$ by an arbitrary function $f(r)$. Then the generalization of the
Kerr-Newman metric takes the form,
\begin{eqnarray}
ds^2 &=&-\frac{f(r) }{\rho ^2}\left( dt- a \sin^2 \theta~ d\phi
\right) ^2+\frac{\rho ^2}{ f(r) }dr^2  \nonumber \\
&&\ + \rho ^2 d\theta ^2+\frac{ \sin
^2\theta}{\rho ^2}\left( a dt-(r^2+a^2) d\phi \right) ^2,
\label{kerrNf}
\end{eqnarray}
The horizon is defined by the surface $f(r=R)=0$. This horizon is generated by the killing vector $\xi^{\mu}$ and the surface gravity $\kappa$ associated with this killing horizon is \cite{MTW},
\begin{eqnarray}
\kappa^2 = - \frac{1}{2} \xi^{\alpha;\beta}\xi_{\alpha;\beta}.
\end{eqnarray}
Using this definition of the surface gravity, it is easy to evaluate the temperature $T$ associated with this horizon as,
\begin{eqnarray}
T = \frac{f'(R)}{4 \pi (R^2 + a^2)} ,
\end{eqnarray}
The Bekenstein-Hawking entropy associated with this horizon is one quarter of the area of the horizon surface. The important thing to note is that unlike spherical geometry the horizon surface here is not simply a $2$-sphere. The area of the horizon can be computed from the $2$-metric on the horizon and it is given by \cite{MTW},
\begin{eqnarray}
\mathcal{A} = 4 \pi (R^2 + a^2).
\end{eqnarray}
Therefore the corresponding entropy associated with this horizon is $S = \pi (R^2 + a^2)$.
 
 With this setting, we need to evaluate the on-horizon components of the Einstein tensor. To do this, we first expand the function $f(r)$ near the horizon at $r = R$ as,
\begin{eqnarray}
f(r) &\approx& f(R) + f'(R)~ (r - R) \nonumber \\
&=& f'(R)~ (r-R),
\end{eqnarray}
and introduce a new coordinate $\eta$ defined by,
\begin{eqnarray}
d\eta = \frac{dr}{\sqrt{f'(R)~ (r-R)}}.
\end{eqnarray}
The horizon is now situated at $\eta = 0$. We first express the metric in terms of these new coordinates $(t , \eta, \theta, \phi)$ to obtain:
\begin{widetext}
\begin{eqnarray}
ds^2 &=& \frac{ \eta^2 \kappa^2 (R^2 + a^2)^2 - a^2 \sin^2(\theta) }{ \rho_{h}^2 + (R^2 + a^2) R \kappa \eta^2} dt^2 + \frac{ 2 a ( H - \eta^2 \kappa^2 (R^2 + a^2)^2 ) \sin^2(\theta) }{ \rho_{h}^2 + (R^2 + a^2) R \kappa \eta^2 } dt d\phi \nonumber \\
&+& ( -\rho_{h}^2 - (R^2 + a^2) R \kappa \eta^2 ) ( d\eta^2 + d\theta^2 ) -
\frac{ H^2 - \eta^2 \kappa^2 a^2 (R^2 + a^2)^2 \sin^2(\theta)}{ \rho_{h}^2 + (R^2 + a^2) R \kappa \eta^2 } \sin^2(\theta) d\phi^2,
\end{eqnarray}
\end{widetext}
where we have defined, $H = (R^2 + a^2 + (R^2 + a^2)R \kappa \eta^2  )$ and $ \rho_{h} = R^2 + a^2 \cos^2(\theta) $. We then evaluate the components of the Einstein tensor for the stationary observers on the event horizon. These observers have the four velocity $ u^{\alpha} = \xi^{\alpha}/\sqrt{- \xi^{\beta}\xi_{\beta}}$. After some straightforward tensor manipulations (for similar calculations, see \cite{medvid} ), we find that the on-horizon ($\eta=0$) components of the Einstein tensor, as seen by the stationary observers, are given by,
\begin{eqnarray}
G^{\hat{t}}_{\hat{t}} = G^{\hat{\eta}}_{\hat{\eta}} = - \frac{R^2 - R f'(R) -a^2}{\rho^4(R, \theta)},
\end{eqnarray}
where the {\it{hat}} over the coordinates implies that the components are being calculated with respect to the stationary observers with four velocity $ u^{\alpha} = \xi^{\alpha}/\sqrt{- \xi^{\beta}\xi_{\beta}}$.

Using this, the $(^{\hat{t}}_{\hat{t}})$ component of the Einstein equations can be written as,
\begin{eqnarray}
G^{{\hat{t}}}_{{\hat{t}}} = 8 \pi T^{{\hat{t}}}_{{\hat{t}}} = 8 \pi T^{{\hat{\eta}}}_{{\hat{\eta}}},
\end{eqnarray}
where the second equality is obtained from the fact that on the horizon, $G^{\hat{t}}_{\hat{t}} = G^{\hat{\eta}}_{\hat{\eta}}$. Therefore, we finally obtain,
\begin{eqnarray}
\frac{R^2 - R f'(R) -a^2}{\rho^4(R, \theta)} = - 8 \pi T^{{\hat{\eta}}}_{{\hat{\eta}}}, \label{eqH}
\end{eqnarray}
Where $T^{{\hat{\eta}}}_{{\hat{\eta}}}$ is the $(^{\hat{\eta}}_{\hat{\eta}})$ component of the energy-momentum tensor on the horizon, given by \cite{MTW};
\begin{eqnarray}
T^{{\hat{\eta}}}_{{\hat{\eta}}} = -\frac{Q^2}{8 \pi \rho^4(R , \theta)}.
\end{eqnarray}
Substituting this in Eq.~(\ref{eqH}), and multiplying both sides by $dR$ we obtain after some straightforward algebra:
\begin{eqnarray}
\frac{f'(R)}{4 \pi(R^2 + a^2)} d(\pi (R^2 &+& a^2)) - \frac{1}{2} \left(\frac{R^2 - a^2}{R^2 + a^2}\right)~dR \nonumber \\
&&\ = -\frac{Q^2}{2(R^2 + a^2)} dR, \label{eqH2}
\end{eqnarray}
In deriving this equation, we have assumed that the radial coordinate of the horizon is changed keeping $a$ as constant. For Kerr-Newman case, $a=J/M$, where $J$ is the angular momentum of the hole. Hence, for simplicity, we have considered a particular variation in $R$, for which, $dJ = a~dM$. Note that, the first term in the left hand side of the above equation is already of the form $T dS$. In order to interpret the rest of the terms we note that, for the Kerr-Newman case, $R =M + \sqrt{M^2 - a^2 - Q^2}$. Using this, Eq.~(\ref{eqH2}) can be witten as,
\begin{eqnarray}
TdS - dM +  \frac{a}{R^2 + a^2} dJ + \frac{R Q}{R^2 + a^2} dQ  = 0.
\end{eqnarray}
It is easy to identify the third term as $\Omega_H dJ$ and the last term as $ \Phi_H dQ$, where $\Phi_H$ is the electrostatic potential on the horizon given by \cite{MTW},
\begin{eqnarray}
\Phi = - A_{\mu} \xi^{\mu} |_{H} =  \frac{R Q}{R^2 + a^2}.
\end{eqnarray}
With these identifications, on the horizon Eq.~(\ref{eqH}) becomes identical to the thermodynamic relation,
\begin{eqnarray}
TdS = dM - \Omega_H dJ - \Phi dQ, \label{thermo}
\end{eqnarray}
which is \textit{identical to the exact form} of the first law of thermodynamics for the Kerr-Newman black hole \cite{kerrthermo,hawking}. It is also easy to see that putting $Q=0$, one can obtain the first law of thermodynamics for pure Kerr horizon. Hence, just like the simple spherically symmetric case, the field equations for the Kerr-Newman spacetime also emerge as a thermodynamic identity under the virtual displacement of the horizon.
\section{Discussion}

In order to interpret these result, we recall that in the standard thermodynamics, the first law provides a connection between two quasi-static equilibrium states, differing infinitesimally in extensive variables like entropy, energy etc., while having the same values for the intensive variables like temperature. In a similar way, Eq.~(\ref{thermo}) can be interpreted as the connection between two quasi-static equilibrium states, where both are the solutions of the Einstein's equations, differing only in the parameters characterizing the horizon. This shows that the gravitational dynamics in Einstein's theory admits a thermodynamic description, \textit{even in the cases} like stationary axis-symmetric spacetime or time dependent evolving horizons thereby broadening the thermodynamic framework for gravity in to more general setting.

We could summarize the broader picture as follows; presence of the causal horizon is a unique feature of gravity because only gravity can effect the causal structure of the spacetime described by light cones. The dynamics of gravity allow a thermodynamic prescription that makes gravity essentially holographic and the dynamical equations involving the metric components can be written as a thermodynamic identity even without the restriction of spherical symmetry. It may be possible to invert this logic and argue that the thermodynamic interpretation of gravity is generic, which itself may be a consequence of the underlying microscopic (\textit{statistical mechanics}) quantum theory. Such an interpretation offers a new outlook towards the dynamics of gravity and therefore is expected to provide valuable clues for any future theory of quantum gravity.

DK and SS thank Gaurang Mahajan for helpful discussions. They are supported by the Council of Scientific \& Industrial Research, India.

\end{document}